\documentclass[aps,showpacs,graphicx]{revtex4}
\usepackage[dvips]{graphicx}
\begin{document}
\title{Properties of  Many-polaron in Fractional Dimension Space }
\author{K. M. Mohapatra}
\affiliation{Department of Physics, Nalanda Institute of Technology, 
Bhubaneswar, Odisha, India.}
\author{B. K. Panda}
\affiliation{Department of Physics, Ravenshaw University, 
Cuttack-753003, Odisha, India.}
\begin{abstract}
Polaron binding energy and effective mass are calculated in the fractional-dimensional 
space approach using the second-order perturbation theory. The effect of carrier
density on the static screening correction of the electron-phonon interaction is 
calculated using the Hubbard's local field factor. It is found that the effective mass 
and the binding energy both decrease with increase in doping.
\end{abstract}
\maketitle
\section{Introduction}
When an electron in the bottom of the conduction band of a polar semiconductor moves, 
its Coulomb field displaces the positive and negative ions with 
respect to each other producing polarization field. The electron 
with the associated phonon cloud is known as a polaron\cite{Mahan}. 
The Fr\"ohlich polaron results when the interaction between electron 
and longitudinal optic 
(LO) phonon is described in the Fr\"ohlich form characterized by the 
dimensionless coupling constant $\alpha$\cite{Frohlich}. 
The polaron in an undoped polar
material calculated with the unscreened electron-phonon interaction is 
termed as a single-polaron. In the weak coupling limit $(\alpha<1)$,
the perturbation theory is applied to calculate ground state energies and 
effective masses of the single-polaron in three-dimensional 
(3D) polar materials\cite{Mahan}, two-dimensional (2D) quantum well (QW) 
structures\cite{Devreese} and one-dimensional(1D) quantum wire structures\cite{Hellman}.  

When the width of the QW is extremely narrow and 
the barrier potential that causes
the in-plane confinement is infinite, the system is purely 2D. The dimension 
increases monotonically with the increase of the well width and the infinitely
wide well exhibits the three-dimensional (3D) behavior.
In a finite QW with narrow well width, the electron envelope function
spreads into the barrier region partially restoring  the 3D
characteristics of the system. The properties of the QW are determined by the 
parameters of the barrier materials. In the QW with large well width, 
the properties of the polaron are calculated taking the bulk values of the 
well material. This has been demonstrated in the
calculation of exciton binding energy\cite{Exciton} and polaron properties 
as a function of well width\cite{Peeters}.
Consequently, the QW with finite well width and barrier height shows
fractional dimensional behavior with the dimension 
$\beta$ lying between 2 and 3\cite{Matos1}. 

The anisotropic interactions in an anisotropic solid
are treated as ones in an isotropic
fractional dimensional space, where the dimension is
determined by the degree of anisotropy\cite{He}.
Thus only a single parameter known as the degree of
dimensionality $\beta$ is needed to describe the system.
The fractional dimensional space is not a vector space and
the coordinates in this space are termed as
{\em pseudocoordinates}\cite{Stillinger}.

   The single-polaron binding energy and effective mass have been derived 
in rectangular and parabolic QWs\cite{Matos2}. When the carrier 
density is high, 
the electron-phonon interaction is dynamically 
screened by the frequency-dependent 
dielectric function and the resulting polaron is termed as the 
many-polaron.  However, for systems with much larger plasma frequency than the 
LO phonon frequency, the static dielectric function can also be a good 
approximation. This happens when the carrier density is very large in the 
system. The properties of many-polaron in the doped ZnS was studied 
by da Costa and Studart\cite{Studart} including the exchange-correlation 
effects beyond random phase approximation. They compared their results by including 
the Thomas-Fermi, Hubbard and more accurate static Sl\"olander-Land-Tosi-Singwi
(SLTS) local field factors. They found that the ground state energies and effective 
masses obtained with the Hubbard's local field correction are similar to those obtained in the SLTS method.  We have therefore implemented the Hubbard's local field factor for calculating properties of the many-polaron.

\section{Polaron Binding Energy and effective mass }

In the second-order  perturbation method, the polaron binding energy 
is calculated from the electron self-energy due to 
electron-phonon interaction as\cite{Mahan}
\begin{equation}
E_{\beta D}=-\Sigma_{\beta D}({\bf k},\xi_{k})|_{{\bf k}=0},
\end{equation}
where $\xi_{k}$ is the electron energy with parabolic band dispersion 
$(\hbar^2k^2/2m_{b})$.  
The effective mass in the same method is defined as
\begin{equation}
\frac{m_{b}}{m^{\ast}}=1+\biggl(\frac{m_{b}}{\hbar^2}\biggr)
\frac{\partial^{2}\Sigma_{\beta D}({\bf k},\xi_{k})}{\partial k^2}\Biggr|_{{\bf k}=0}
\end{equation}
where $m_{b}$ is the band electron mass.  
The leading-order contribution to the electron self-energy due to the 
electron-phonon interaction at zero temperature is given by
\begin{equation}
\Sigma_{\beta}(k,\xi_{\bf k})=\sum_{q}\frac{|M_{\beta}(q)|^{2}}
{\epsilon^{2}_{\beta }(q,0)}
\left[\frac{1}
{\xi_{\bf k}-\xi_{{\bf k}+{\bf q}}+\hbar\omega_{LO}}\right],
\end{equation}
where $\omega_{LO}$ is the LO phonon frequency.
Here$M_{\beta D}$ is the electron-phonon interaction in the Fr\"ohlich form 
is given as
\begin{equation}
M_{\beta D}(q)=-i\hbar\omega_{LO}\Biggl(\frac{(4\pi)^{\frac{\beta-1}{2}}\Gamma\biggl(\frac{\beta-1}{2}\biggr)R_{p}\alpha}
{q^{\beta-1}\Omega_{\beta }}\Biggr)^{\frac{1}{2}},
\end{equation}
where $\Gamma$ is the Euler-gamma function and
$R_p=\sqrt{\hbar/2m_b\omega_{LO}}$ is the polaron radius. 
The dimensionless coupling constant $\alpha$ is defined as
\begin{equation}
\alpha=\frac{e^2}{2\hbar\omega_{LO}R_{p}}
\biggl(\frac{1}{\epsilon_{\infty}}-\frac{1}{\epsilon_{0}}\biggr),
\end{equation}
where $\epsilon_{0}$ and $\epsilon_{\infty}$ are the static and high-frequency dielectric constants, respectively.  
The static dielectric function including the local-field factor is defined as
\begin{equation}
\epsilon_{\beta D}(q,rs)=\epsilon_{\infty}\Biggl[\frac{1-[1-G_{\beta D}(q,rs)]
V_{\beta D}(q)\chi_{\beta D}(q,rs)}
{1+G_{\beta D}(q,rs)V_{\beta }(q)\xi_{\beta D}(q,rs)}\Biggr],
\end{equation}
where $G_{\beta D}(q,rs)$ is the Hubbard local-field-factor given by

\begin{equation}
G_{\beta D}(q,rs)=\frac{1}{2}\frac{q^{\beta-1}}{(q^2+k^{2}_{F})^{\frac{\beta-1}{2}}}.
\end{equation}
The dimensionless density parameter $rs$ is given by
\begin{equation}
k_{F}r_{s}a_{B}=\biggl[2^{\beta-1}\Gamma^{2}
\biggl(1+\frac{\beta}{2}\biggr)\biggr]^{\frac{1}{\beta}}
\end{equation}
 where $a_{B}$ is Bohr atomic radius and $k_{F}$ is Fermi wave vector.

The irreducible polarizability function 
 $\xi_{\beta D}(q,rs)$ is defined as
\begin{equation}
\chi^{0}_{\beta D}(q,r_{s})=\sum_{{\bf k}}\frac{n_{F}({\bf k}+{\bf q})-
n_{F}({\bf k})}{\xi_{{\bf k}+{\bf q}}-\xi_{\bf k}}.
\end{equation}
In the fractional dimensional method the sum over {\bf k} is 
transferred to integrating as 
\begin{equation}
\sum_{\bf k}=\int^{\infty}_{0}k^{\beta -1}DJ\int^{\pi}_{0}\sin^{\beta-1}\theta
d\theta
\end{equation}
Using Eq.(10) in Eq. (9), we find
\begin{equation}
\chi^{0}_{\beta D}(q,rs)=-\frac{2^{3-\beta}m_{b}k^{\beta}_{F}}
{\pi^{\frac{\beta-1}{2}}\hbar^2\beta q^{2}\Gamma\biggl(\frac{\beta-1}{2}\biggr)} \int^{\pi}_{0}{_2F_1}\biggl(1,\frac{\beta}{2};
\frac{\beta+2}{2};\frac{4 k^{2}_{F}\cos^{2}\theta}{q^2}\biggr)\sin^{\beta-2}\theta d\theta,
\end{equation}
where $_{2}F_{1}$ is the Gauss hypergeometric function.

The Fourier transform of $e^2/r$ in the fractional dimensional space 
is obtained as
\begin{equation}
V_{\beta D}(q)=\frac{(4\pi)^{\frac{\beta-1}{2}}e^2
\Gamma\biggl(\frac{\beta-1}{2}\biggr)}
{q^{\beta-1}}\label{eq:Coul}
\end{equation}

Using Eqs. (6), (7), (10) and (12) in Eq. (1), the binding energy is
obtained as
\begin{equation}
E_{\beta D}=-\alpha\hbar\omega_{LO}\frac{\Gamma\biggl(\frac{\beta}{2}\biggr)}
{\sqrt{\pi}\Gamma\biggl(\frac{\beta}{2}\biggr)}
\int^{\infty}_{0}\frac{dq}{\epsilon_{\beta}(q,r_{s})(q^2+1)}.
\end{equation}

 Similarly the effective mass in Eq.(2) can be obtained as
\begin{equation}
\frac{m_{b}}{m^{\ast}}=1-4\alpha
\frac{\Gamma\biggl(\frac{\beta-1}{2}\biggr)}
{\sqrt{\pi}\beta\Gamma\biggl(\frac{\beta}{2}\biggr)}
\int^{\infty}_{0}\frac{q^2dq}{\epsilon^{2}_{\beta D}(q,rs)(q^2+1)^{3}}.
\end{equation}

For nondegenerate systems$(rs\rightarrow \infty)$, $\epsilon_{\beta D}=1$. The integrals in above Eqs(11) and (12) can be analytically evaluated.Now
The binding energy s derived as
\begin{equation}
E_{\beta D}=-\frac{1}{2}\alpha\hbar\omega_{LO}
\frac{\sqrt{\pi}\Gamma\biggl(\frac{\beta-1}{2}\biggr)}
{\Gamma\biggl(\frac{\beta}{2}\biggr)}
\end{equation}

and the effective mass is given by
\begin{equation}
\frac{m^{\ast}}{m_{b}}=1-\frac{1}{4}\alpha
\frac{\sqrt{\pi}\Gamma\biggl(\frac{\beta-1}{2}\biggr)}
{\beta\Gamma\biggl(\frac{\beta}{2}\biggr)},
\end{equation}
\section{Results and  Discussions}

we have taken the several parameters of GaAs to calculate 
polaron properties.The band mass  $m_b$=0.067$m_{0}$, where $m_{0}$ 
is the electron mass,   
$\epsilon_{\infty}=13.18$ and $\epsilon_{0}=10.89$. The value of 
$\alpha$=0.03 which is appropriate for calculating polaron properties 
in the weak coupling limit. The LO phonon energy $(\hbar\omega_{LO})$ 
is taken as 36.25 meV.

The binding energy and effective masses calculated for several $r_{s}$ values 
for dimensions $\beta$=2, 2.5 and 3 are shown in Table I. Both binding 
energies and effective masses are found to increase with the increasing 
$r_{s}$. This suggests that the polaron properties decrease with the 
increasing carrier density. This results due to screening of the 
electron-phonon interaction. 

The polaron properties also decrease as the dimension decreases. As the 
dimension of the system decreases, the system becomes more confined. 
The confinement of the system decreases the physical properties. 

Although the static Hubbard's local-field-factor for the screening of the 
electron-phonon interaction correctly predicts  
$r_{s}$ and $\beta$ dependence of the physical properties, it does not  
include the exchange and correlation effects properly. It is much 
higher than the dynamic 
local field factor as the later correctly includes  
the exchange and correlation effects. Therefore it is required that 
we calculate the dynamic local field factor using quantum version of the 
STLS method and screen the electron-phonon interaction term. The Matsubara 
frequency summation method is the right direction for achieving it. 
Such a work is in progress in our group.  

\vfill
\eject

\begin{table}
\begin{ruledtabular}
\begin{tabular}{ccccccc}
&\multicolumn{3}{c}{Polaron Energy (eV)}&\multicolumn{3}{c}{Effective mass}\\
\hline
$r_{s}$ & $\beta=2$ & $\beta=2.5$  & $\beta=3$ & $\beta=2$ & $\beta=2.5$ & 
$\beta$=3 \\
0.001 & 0.416 & 0.296 & 0.210 & 1.120 & 1.067 & 1.041 \\
0.01 & 0.417 & 0.315 & 0.255 & 1.121 & 1.070 & 1.047 \\
0.10 & 0.424 & 0.325 & 0.272 & 1.123 & 1.071 & 1.049 \\
1.0 & 0.430 & 0.328 & 0.274 & 1.124 & 1.072 & 1.050 \\
10.0 & 0.430 & 0.340 & 0.275 & 1.125 & 1.073 & 1.051 \\
\end{tabular}
\end{ruledtabular}
\end{table}
Table I: Polaron binding energy shift and effective mass as a function 
of $r_{s}$ for dimensions $\beta$=2, 2.5 and 3. 

\section{Conclusions}

In the present work the binding energy shift and effective mass of the 
many-polaron system are calculated in the fractional dimensional method
by screening the electron-phonon interaction term. The exchange-correlation 
effects beyond the random phase approximation method is included in the 
Hubbard's method. The polaron properties are found to decrease with the 
carrier density.

\end{document}